\documentstyle[twoside,fleqn,espcrc2,epsf]{article}

\newcommand{\AmS}{{\protect\the\textfont2
  A\kern-.1667em\lower.5ex\hbox{M}\kern-.125emS}}

\newcommand{\bd}{\begin{displaymath}}
\newcommand{\ed}{\end{displaymath}}
\newcommand{\be}{\begin{equation}}
\newcommand{\ee}{\end{equation}}
\newcommand{\bea}{\begin{eqnarray}} 
\newcommand{\eea}{\end{eqnarray}}
\newcommand{\bt}{\begin{tabular}}
\newcommand{\et}{\end{tabular}\newline}
\newcommand{\x}{\mbox{x}}

\newcommand{\Pn}{\psi}
\newcommand{\Pb}{\bar{\psi}}

\newcommand{\la}{\langle}
\newcommand{\ra}{\rangle}

\title{
\vspace{-3.65cm}                                            
{\normalsize DESY 99--116}    \\[-0.2cm]                    
{\normalsize HUB--EP--99/39}  \\[-0.2cm]                    
{\normalsize August 1999} \\                                
\vspace{2.25cm}                                             
Higher-twist contributions to the Structure Functions coming from
4-fermion operators\thanks{Talk
given by S. Capitani at Lattice 99,                         
               Pisa (Italy).}}                              

\author{
S.~Capitani\address{Deutsches Elektronen-Synchrotron DESY, Notkestrasse 85, 
D-22607 Hamburg, Germany}, 
M.~G\"ockeler\address{Universit\"at Regensburg, Institut f\"ur Theoretische 
Physik, D-93040 Regensburg, Germany}, 
R.~Horsley\address{Humboldt-Universit\"at, Institut f\"ur Physik, 
Invalidenstrasse 110, D-10115 Berlin, Germany}, 
B.~Klaus\address{Deutsches Elektronen-Synchrotron DESY and 
NIC, Platanenallee 6, D-15738 Zeuthen, Germany}$^{\rm ,e}$,
V.~Linke\address{Freie Universit\"at Berlin, Arnimallee 14, 
D--14195 Berlin, Germany},
P.~Rakow$^{\rm b}$, 
A.~Sch\"afer$^{\rm b}$
and G.~Schierholz$^{\rm a,d}$
}

\begin{document}

\begin{abstract}
We evaluate the contribution of a class of higher--twist
operators to the lowest moment of the Structure Functions,
by computing appropriate matrix elements of six four--fermion
operators in the quenched approximation. Their perturbative
renormalization constants and mixing coefficients are calculated
in the 't Hooft--Veltman scheme of dimensional regularization,
using codes written in the algebraic manipulation computer
language FORM.
\end{abstract}

\maketitle

\section{INTRODUCTION}

By performing analytic calculations and numerical simulations with 
quenched Wilson fermions, we have computed the contributions of some 
classes of higher--twist operators to moments of the structure functions.

These moments are related via an OPE to hadron matrix 
elements of local operators, whose twist is the difference between their 
dimensions and their spin. Twist--2 operators give the leading contributions, 
while twist--4 operators correspond to the $1/Q^2$ power corrections:
\bea
\noalign{$\displaystyle M_n (Q^2) = \int_0^1 {\rm d}\x\,\x^{n-2}F_2(\x,Q^2)$}
  &&= C^{(2)}_n(Q^2/\mu^2,g(\mu)) \, A^{(2)}_n(\mu) \label{eq:mom} \\
  &&+ C^{(4)}_n (Q^2/\mu^2,g(\mu)) \, \frac{A^{(4)}_n (\mu)}{Q^2} 
  + O\left(\frac{1}{(Q^2)^2}\right) \nonumber .
\eea
The leading twist contribution can be written as 
$M_n^{(2)}=\sum_f Q_f^2 \, \la\x_f^{n-1}\ra$, where $Q_f$ 
is the charge of the quark of flavor $f$. At twist--4 level we consider the 
$I=2$ pion structure function~\cite{gottliebmorelli}
\be
F_2^{I=2} = F_2^{\pi^+}+F_2^{\pi^-}-2F_2^{\pi^0} ,
\label{eq:I2}
\ee
which belongs to a flavor $27$--plet, receives contributions only from 
4--fermion operators and therefore cannot mix with twist--2 
operators~\footnote{When mixing with operators of leading twist is forbidden, 
one gets rid of renormalon ambiguities completely. When such a mixing 
occurs, a non--perturbative computation of the Wilson coefficients 
in the way proposed in~\cite{cf1,cf2} would avoid renormalon problems. 
That method 
will be a useful complement to the present one for an understanding of 
higher--twist effects. So far however only 2--quark states have been 
simulated.}. 
While the Wilson coefficients $C^{(k)}_n$ can be calculated in perturbative 
QCD~\cite{js,ht}, the computation of the matrix elements $A^{(k)}_n$ is a 
strong interaction problem that cannot be treated perturbatively. 
Some results have been obtained in the past using the MIT bag model~\cite{js}, 
but a calculation in a reliable model--independent way and from first 
principles is only possible in lattice QCD.

The leading--twist contribution to the lowest unpolarized moment is 
given by the operator
\be
  O_{\mu_1\mu_2}=\frac{\rm i}{2}\Pb\gamma_{\mu_1}
  \stackrel{\leftrightarrow}{D}_{\mu_2} \Pn - \mbox{traces} ,
\ee
with Wilson coefficient $C_2^{(2)} = 1+O(g^2)$ and forward matrix elements 
$\langle\vec{p}|O_{\{\mu_1\mu_2\}}|\vec{p}\rangle = 2
A_2^{(2)} [p_{\mu_1}p_{\mu_2} -\mbox{traces}]$. The twist--4 contribution 
involves various operators~\footnote{For discussions of twist--4 operator 
bases and general higher--twist effects, see also~\cite{ht,cf1,lat98}.}, 
but only 4--fermion operators can contribute to the $I=2$ 
combination~(\ref{eq:I2}), and the only 4--fermion operator that appears 
in~(\ref{eq:mom}) at order $g^2$ is
\be
  A^c_{\mu_1\mu_2}= \Pb\gamma_{\mu_1}\gamma_5
  t^a\Pn\, \, \Pb\gamma_{\mu_2}\gamma_5 t^a\Pn - \mbox{traces} ,
\ee
with Wilson coefficient $C_2^{(4)} = g^2 (1+O(g^2))$ and forward matrix 
elements $\langle\vec{p}|A^c_{\{\mu_1\mu_2\}}|\vec{p}\rangle = 2
A_2^{(4)} [p_{\mu_1}p_{\mu_2} -\mbox{traces}]$.
This operator mixes under renormalization, and on the lattice there is
even more freedom to mix, given the lower symmetry of the theory compared 
to the continuum.

\section{PERTURBATIVE RENORMALIZATION}

We have computed pion matrix elements of six 4--fermion operators and 
perturbatively calculated their renormalization factors, which relate the 
lattice numbers to physical quantities in the $\overline{\rm{MS}}$ scheme, 
using the 't Hooft--Veltman prescription for $\gamma_5$. We consider the 
following operators in Euclidean space, symmetrized in $\mu$ and $\nu$:
\bea
    V^c_{\mu\nu}  &=& \Pb \gamma_{\mu} t^a \Pn 
    \, \, \Pb \gamma_{\nu} t^a \Pn -\mbox{traces}\\
    A^c_{\mu\nu}  &=& \Pb \gamma_{\mu} 
    \gamma_5 t^a \Pn \,  \, \Pb \gamma_{\nu} \gamma_5 t^a \Pn-\mbox{traces} \\
    T^c_{\mu\nu} &=& \Pb \sigma_{\mu\rho} 
    t^a \Pn \, \,  \Pb \sigma_{\rho\nu} t^a \Pn-\mbox{traces}\\ 
    V_{\mu\nu} &=& \Pb \gamma_{\mu} \Pn 
    \, \,  \Pb \gamma_{\nu} \Pn -\mbox{traces}\\
    A_{\mu\nu} &=& \Pb \gamma_{\mu} 
    \gamma_5 \Pn \, \,  \Pb \gamma_{\nu} \gamma_5 \Pn -\mbox{traces}\\
    T_{\mu\nu} &=& \Pb \sigma_{\mu\rho} 
    \Pn \, \,  \Pb \sigma_{\rho\nu} \Pn-\mbox{traces} .
\eea
Not all above operators are present in the OPE~(\ref{eq:mom}), 
they appear however when one computes the lattice radiative 
corrections~\footnote{In principle also gauge--variant operators could appear 
in the mixing, but there are no such 4--fermion operators with dimension 6, 
and 2-fermion operators do not contribute.}.

One needs the renormalization factors of the operators on the lattice 
as well as in the continuum, in order to obtain physical continuum 
matrix elements from the Monte Carlo simulations. The connection between 
the two is
\be
\langle O_i^{\rm cont} \rangle = \sum_j \Big( \delta_{ij} - 
\frac{g^2}{16 \pi^2} \Delta R_{ij} \Big) 
\cdot \langle O_j^{\rm lat} \rangle ,
\ee
where
\be
\langle O_i^{\rm cont,lat} \rangle = \sum_j \Big( \delta_{ij} 
+ \frac{g^2}{16 \pi^2} R_{ij}^{\rm cont,lat} \Big) 
\cdot \langle O_j^{\rm tree} \rangle
\ee
are the renormalized (in our case in the $\overline{\rm{MS}}$ scheme) 
continuum and unrenormalized lattice \linebreak 1--loop expressions. 
We computed the matrix \linebreak elements on quark states, in a general 
covariant gauge.
The differences $\Delta R_{ij} = R_{ij}^{\rm lat} - R_{ij}^{\rm cont}$ 
enter then in the renormalization factors
\be
Z_{ij} (a\mu,g)=\delta_{ij} -\frac{g^2}{16 \pi^2} \Delta R_{ij},
\ee
which convert from the lattice to the continuum, and are the object of our 
computations.

To evaluate the Feynman diagrams and obtain the algebraic expressions 
for the renormalization factors, we run computer codes written in the 
symbolic manipulation language FORM. To properly deal with the $\gamma_5$ 
matrices we have also 
written additional computer routines which are able to perform computations 
with the 't~Hooft--Veltman prescription. This is the only one proven 
to be consistent for $\gamma_5$, and if we were not using it we would not 
find the right mixing factors between axial, vector and tensor operators.

Implementing the 't~Hooft--Veltman scheme in a computer program is a 
challenging task, compounded by the fact that we are dealing here with lattice 
calculations. In fact, $d$--dimensional sums are split in $4$ and 
$(d-4)$ dimensions, in which the Dirac algebra is different from usual 
(see for example~\cite{gammatrica}). 
The treatment of $\gamma$--matrices, already complicated by the 
non--validity of the summation convention on the lattice (as Lorentz 
invariance is broken), is then subject to much more complicated 
rules. When running the codes, the sum 
splitting increases the number of terms by about one order of magnitude, 
and this slows down considerably the computations, besides generating 
problems in memory management as well. We have found the computations 
presented here, which only concern Wilson fermions without any improvement, 
to be already very demanding.

We use the Kawai method~\cite{kawai} in dealing with divergences,
which allows a strong check in the analytic lattice calculations by using 
two different intermediate regularizations: 
dimensional regularization with the 't Hooft--Veltman prescription on the 
lattice, and an infrared--mass regularization.

Using the 't~Hooft--Veltman prescription for the $\gamma_5$ matrices, 
the one--loop renormalized operators in the $\overline{\rm{MS}}$ scheme 
can be written in terms of the lattice operators (for $\mu=1/a$) 
as~\footnote{For a similar calculation in the context of weak matrix
elements, involving a different set of operators and a different scheme, 
see also~\cite{sharpe}.}
\bea
  (V^c)^R &=& V^c -g^2 \Big( 0.281578 \,\, V^c + 0.000532 \,\, A \nonumber \\
          & & + 0.000997 \,\, A^c  + 0.022899 \,\, T^c\Big) \nonumber \\
  V^R &=& V -g^2 \Big( 0.348170 \,\, V + 0.002393 \,\, A^c \Big) \nonumber \\
  (A^c)^R &=& A^c -g^2 \Big( 0.291756 \,\, A^c + 0.000532 \,\, V \nonumber \\
      & & + 0.000997 \,\, V^c  + 0.006785 \,\, T \nonumber \\
      & & + 0.012722 \,\, T^c \Big) \nonumber \\
  A^R &=& A -g^2 \Big( 0.266750 \,\, A 
                + 0.002393 \,\, V^c  \nonumber \\
      & & + 0.030533 \,\, T^c \Big).
\label{eq:pertres}
\eea
The complete results for general $N_c$ are given in~\cite{big}. Although
used here for the pion, they are of more general application, 
and we will use them also for the nucleon case.

\section{RESULTS}

We made simulations on a $16^3\times 32$ lattice at quenched $\beta=6.0$, 
on a Quadrics QH2 at DESY Zeuthen, with a statistics of 400 configurations.
To extract the expectation values of the operators we computed ratios of
three--point to two--point functions of the pion, 
$R_{O}(t,\tau)= \la\pi(t)O(\tau)\pi(0)\ra / \la \pi(t)\pi(0)\ra$.

After extrapolating the matrix elements for $F_2^{I=2}$ to the 
chiral limit and using the renormalization factors~(\ref{eq:pertres}) 
at $Q^2 = \mu^2 = a^{-2}$, we obtain $A_2^{(4)} = 0.085(27) f_\pi^2$,
which multiplied by $C_2^{(4)}$ and the kinematical factor gives the twist--4 
contribution to the lowest moment of $F_2^{I=2}$:   
\be
  M_2^{(4)} = 0.085(27)\frac{f_\pi^2 g^2}{Q^2}+O(g^4).
\ee
For the leading contribution we have $\la\x_i\ra=0.273(12)$ 
(independent of flavor), so that weighting with the charges, say of 
the $\pi^+$, one gets~\cite{best}
\be
M_2^{(2)\pi^+}=0.152(7)+O(g^2).
\ee
The relative magnitude of the twist--4 and twist--2 contributions to the 
lowest moment of the pion structure functions turns out to be small, 
but is expected to be larger for the nucleon, where the scale 
$f_\pi$ is replaced by $m_N$.
We are now studying the nucleon with the same machinery and using the same 
renormalization factors as here.

\end{document}